\documentclass[twocolumn,showpacs,showkeys,unsortedaddress]{revtex4-1}
\usepackage{graphicx}
\usepackage{amsmath} 
\begin{document}
\renewcommand{\thesubsection}{\arabic{subsection}}

\title{Climate Change Alters Diffusion of Forest Pest: A Model Study}

\author{Woo Seong Jo}
\author{Beom Jun Kim}
\email[Corresponding author, E-mail: ]{beomjun@skku.edu}
\affiliation{Department of Physics, Sungkyunkwan University, Suwon 16419, Republic of Korea}
\author{Hwang-Yong Kim}
\affiliation{International Technology Cooperation Center, Rural Development Administration, Jeonju 54875, Republic of Korea}


\begin{abstract}
Population dynamics with spatial information is applied to understand
the spread of pests. We introduce a model describing how pests spread in 
discrete space. 
The number of pest descendants at each site 
is controlled by local information such as temperature, precipitation, and the density of pine trees. 
Our simulation leads to a pest spreading pattern comparable to the
real data for pine needle gall midge in the past.
We also simulate the model in two different climate conditions 
based on two different representative concentration pathways
 scenarios for the future.
We observe that after an initial stage of a slow spread of pests, a sudden change in the spreading speed occurs, which is soon followed by a large-scale outbreak.
We found that a future climate change causes the outbreak point to occur
earlier and that the detailed spatio-temporal pattern of the spread 
depends on the source position from which the initial pest infection starts.
\end{abstract}

\keywords{Population dynamics, Pest spreading, Lattice model, Climate change, RCP scenarios}
\pacs{89.75.-k, 87.23.Cc, 87.53.Vb}

\maketitle

\section{Introduction}

Population dynamics has been established as one of the successful mathematical
methods for describing the temporal dynamics of populations in physics~\cite{PRE-extinction,
PRE-pattern, PRE-extinction2} and biology~\cite{MB-text, MB-logistic,
Ecology-PD}. The methodology in population dynamics often takes different forms:
Full-mixing models (i.e., mean-field models in physics) with quantities
averaged over whole spatial locations, and models with spatial information
explicitly taken into account (i.e., structured population models) have been
widely used.  The advantage of the mean-field approach is that the equation for population dynamics often
becomes mathematically tractable, and one can clearly understand what happens in
a long-time limit, i.e., species will be extinct or not. However, such
a mathematical tractability comes at a cost: In the real world of nature, 
all species are locally embedded in a large-scale geographic space with finite
dimensions of two or three.  Living agents located in space cannot interact with all 
other agents in a finite time, and their behaviors are spatially and temporarily 
limited. Such a spatial constraint leads to remarkably different results, 
and the solution from the mean-field approach often fails to explain empirical observations.
For this reason, researchers have been trying to integrate spatial 
information into dynamics in the design of their models to mimic the real world
of nature.

Cellular Automata (CA) is often used as a tool for spatially explicit models. 
In the CA approach, space is approximated as a discrete lattice whose
resolution needs to be fine enough to properly describe local information.  
Besides discretized space, the quantities describing the system are assigned
to each of the sites. It is one of differences between the CA and the individual-based model, 
where agents have their own properties regardless of their location.
 In the present work, we use a discrete lattice like CA, but use the population
density as a local variable defined on each lattice point.
The time evolution of the population density is given by a dynamic equation with a transition rate and local interactions as key ingredients similarly to 
Ref.~\onlinecite{BMB-Caswell}.
A model defined in a discrete lattice is very efficient from a the computational point of view 
because much spatial information can be integrated as spatially discrete variables. 
Such approaches have been utilized for various biological systems like 
vegetation dynamics~\cite{CA-vegetation-Harada, CA-vegetation-Hiebeler,
CA-vegetation-Ikegamia}, epidemics~\cite{CA-epidemics-Rhodes,
CA-epidemics-AIDS}, and spread of pests~\cite{pest-newphysics,CA-pest-Brockhurst,
CA-pest-Chon}, as well as variant subjects in physics~\cite{game1, game2, jkps-lattice}.
 
The spread of pests in vegetation has been a critical issue because  a very
long time is needed for ruined vegetation to recover. In Japan and South Korea, the spread
of the pine needle gall midge (PNGM) has been a serious problem from the early
$1900$'s. When a pine tree is parasitized by PNGMs, the infected trees wither
and can eventually die. Ecological studies of the PNGM have been conducted to
identify conditions for the PNGM to spread broadly, and the spreading pattern
of withered forest have also been empirically investigated~\cite{PNGM-1983,
PNGM-1985, PNGM-2007}.  Computational approaches with more available data 
have been developed, and it has become possible to use simulational methods 
for the study of PNGM spread~\cite{CA-pest-Chon}.
A machine-learning technique has been used to forecast the spread of damage~\cite{ANN-PNGM-Chon, ANN-PNGM-Chung}, 
and images from satellites have also been analyzed~\cite{satellite-PNGM}.  Although population
dynamics in a discrete lattice can be a very useful research framework due to its
flexibility and computational efficiency, not many studies, with a few exception, have used this method~\cite{CA-pest-Chon}. 

Accurate prediction of future climate change is very difficult.
Even when the computational framework for the forecast is given,
the forecast results can differ depending on the assumptions the model uses.  
In reality, the number of quantitative assumptions for future environmental conditions
can be huge. If different studies use different assumptions among many possibilities, 
comparing the results of one study with those of other studies is difficult. 
Scholars in the research area of future climate change have agreed on
a few official scenarios. 
The first scenario was introduced by the Intergovernmental Panel on Climate Change (IPCC) 
in $1992$~\cite{IS92}.  In 2000, the IPCC published the second version of a scenario called the 
Special Report on Emissions Scenarios (SRES)~\cite{SRES}.
The SRES have four categories, 
which were discussed in two follow-up reports, i.e., Third Assessment Report (TAR)
and Assessment Report Four (AR4).  Each category shows the amount of emission
of greenhouse gas, which is determined by the speed of development, human
population, and other possible causes for generating gas.  The latest scenario
of climate change is called Representative Concentration Pathways
(RCPs)~\cite{IPCC2014}, which is an improved version of the SRES with the
latest climate data in $2014$.  RCPs contain four categories called
RCP2.6, RCP4.5, RCP6.0, and RCP8.5.  The first category, RCP2.6 assumes
that the greenhouse effect can be reduced by nature itself, which is doubtful
right now.  As the number in each category increases from 2.6 to 8.5, 
the emission of greenhouse gas is assumed to increase more in the future. 
The RCP8.5 scenario assumes that the concentration of greenhouse gas will
follow the current trend.  Most studies on climate change use one
of the RCP scenarios with different greenhouse gas concentrations for 
the future and try to predict local climate variables such as 
temperature and precipitation.

In this paper, we introduce a model to describe the temporal dynamics and the spread of
a population of PNGMs in the two-dimensional lattice model of South Korea. For
consistency with real field data, small initial PNGM densities are assigned
to three cities, Incheon, Mokpo and Busan, and the spread patterns are
predicted by using climate scenarios RCP4.5 and RCP8.5.  We observe
that the patterns of spread are comparable with the field data observed in the
past.  Furthermore, we also investigate how different climate scenarios 
affect the future prediction of the spread pattern of PNGM.

\section{Method}
\label{sec:method}

\subsection{Climate Conditions}
Future climate changes in South Korea have been estimated
based on the RCP scenarios. Each such estimate, 
RegCM4~\cite{RegCM4}, SNURCM~\cite{SNURCM},
GRIMs~\cite{GRIMs}, WRF~\cite{WRF}, and HadGEM3-RA~\cite{HadGEM3}, 
has limitations of its own and contains inevitable uncertainty 
about initial conditions. Later,
MME5s~\cite{MME5s}, which uses the concept 
of ensemble, combining estimates from five climate predictions,
was suggested in order to overcome such limitations.
In the present study, we use the future climate estimates of
the MME5s in Climate Information Portal~\cite{CIP}. 
The climate data from MME5s have a fine spatial resolution of 1km $\times$ 1km 
for every month from $2021$ to $2050$.
We choose climate data based on two RCP scenarios, RCP4.5 and RCP8.5, 
to compare our results based on different assumptions for the future
concentration of greenhouse gas.

\subsection{Model}
\label{subsec:model}
Once the spatial climate conditions are fixed from the data based on MME5s with
RCP4.5 or RCP8.5, we then apply our structured CA model for the spread of
the PNGM. Heretofore, we will simply call a PNGM as a midge.
Let $\rho_t({\bf r})$ be the density of adult midges  at  a  discrete lattice point
${\bf r}$ and at a discrete time $t$ (we fix the unit of time as unity, which
corresponds to one year, the life cycle of the PNGM).
The adult midge density is modeled to evolve in time as
\begin{equation}
\label{eq:rho_tr}
{\rho}_{t} ({\bf r}) = \lambda({\bf r}; T, P, \psi) 
\overline{\rho}_t({\bf r}) 
\exp \left[-\frac{ \overline{\rho}_t({\bf r}) } {\psi_t({\bf r})}\right] ,
\end{equation}
where $\overline{\rho}_t({\bf r})$ is the density of midge eggs to be explained below
and $\lambda({\bf r})$ is the position-dependent survival rate, which depends on
the local values of the temperature $T$, the precipitation $P$, and the tree
density $\psi$.  The previous generation of adult midges at $t-1$ cannot survive
more than a year; thus, the adult midges at $t$ are all born from midge eggs left
by adult midges in the previous generation at $t-1$. Accordingly, the midge density $\rho_t$
at $t$ does not directly depend on $\rho_{t-1}$, but it depends on the density
$\overline{\rho}_t$ of midge eggs left by the previous generation of
adult midges at time $t-1$.  When $\overline{\rho}_t({\bf r}) /
\psi_t({\bf r})$ is small, $\exp [-\overline{\rho}_t({\bf r}) /
\psi_t({\bf r})] \approx 1-\overline{\rho}_t({\bf r}) / \psi_t({\bf r})$, and
the right-hand side of Eq.~(\ref{eq:rho_tr}) takes the form of
the corresponding term in the standard logistic equation. 
In the original logistic equation, the population growth is controlled by the
carrying capacity. The population can exceed the carrying capacity in the
logistic equation, but then the growth rate becomes negative, reducing 
the population afterward. However, our growth model governed by 
Eq.~(\ref{eq:rho_tr}) is somewhat different. If we replace the exponential
term by the standard form in the original logistic equation, the midge
density can be negative, causing our growth model to fail. 
In order to avoid this catastrophe in evolving dynamics, we have, thus,
introduced the form $xe^{-x}$, instead of the quadratic form in the 
conventional logistic equation. 
The exponential form is equivalent to the standard logistic form $x(1-x)$
when $x$ becomes smalls and prevents the midge density from being
negative when the egg density exceeds the tree density. 
Therefore, the advantage of our growth model~(\ref{eq:rho_tr}) is twofold:
it is consistent with the conventional logistic equation when the egg density
is small, and it also controls the unwanted catastrophic failure in the model.
One can recognize then that the tree density $\psi$
plays the role of the carrying capacity of midge eggs, which appears to be a
reasonable approximation since adult midges lay eggs in pine trees. 
From this reasoning, one can see that we normalize the egg density $\bar\rho$ with
respect to the tree density $\psi$ in such a way that the maximum 
egg density is defined to be proportional to the tree density. 
The proportionality constant can be absorbed into the definition of
the growth rate $\lambda$, giving 
us Eq.~(\ref{eq:rho_tr}).

The position-dependent survival rate $\lambda({\bf r})$ in
Eq.~(\ref{eq:rho_tr}) is affected by local climate conditions
like the temperature and the precipitation.
In our model, one unit of time corresponds to one year; 
thus, the temperature $T$ and the precipitation $P$
need to be defined in some average sense for each year.
We note that most midges become adults from pupa and lay eggs in June; 
thus, we use the time-averaged temperature in June as $T$. 
In existing literature on the PNGM~\cite{ANN-PNGM-Chon}, 
the moisture of soil has been shown to be one of the crucial factors for a midge
to grow into an adult. We, thus, calculate the average precipitation 
from March to May and use it for $P$.  
The survival of the midge larvae to become an adult midge is
known to depend on strongly the temperature~\cite{PNGM-2007}: 
Survival probability shows a slightly skewed bell shape
for the temperature range between $12^{\circ}\text{C}$ and $30^{\circ}\text{C}$. 
We thus use such suitable climate conditions for the PNGM to
grow to adults and write the survival rate $\lambda({\bf r})$ as
\begin{widetext} 
\begin{equation}
\label{eq:lambda}
\lambda({\bf r}; T, P, \psi)  = A \psi_t({\bf r}) [T_t({\bf r}) - T_{\rm min}] 
[T_{\rm max} - T_t({\bf r}) ] G[P_t({\bf r})],
\end{equation}
where $A$ is the normalization constant to make $\lambda$ in the interval $[0,1]$ and $P
({\bf r})$ is the above mentioned time-averaged
precipitation at ${\bf r}$. To mimic the suitable temperature range, 
we have chosen in Eq.~(\ref{eq:lambda}) a concave quadratic form to approximate the bell-shaped curve reported in 
Ref.~\onlinecite{PNGM-2007} with $T_{\rm min} = 12^{\circ}\text{C}$ 
and $T_{\rm max} = 30^{\circ}\text{C}$. 
Although more rain has been reported to be better in 
Ref.~\onlinecite{ANN-PNGM-Chon}, we suppose that the marginal gain must be 
very small if the precipitation is too large. 
As a rough approximation for such a dependence on $P$, we write 
\begin{equation}
\label{eq:G}
G[P_t({\bf r})] = \begin{cases}
 0,  & \text{for } P_t({\bf r}) < P_{\rm min}, \\ 
 [P_t({\bf r}) - P_{\rm min}] / (P_{\rm max} - P_{\rm min} ),  & \text{for } P_{\rm min} \leq P_t({\bf r}) \leq P_{\rm max}, \\ 
 1,      & \text{for } P_t({\bf r}) > P_{\rm max}, 
\end{cases}
\end{equation}
\end{widetext}
where $P_{\rm min} = 20$mm and $P_{\rm max} = 100$mm are suitably chosen based on 
the average precipitation from March to May in Gyeonggi province in Korea. 
Since $G(P)$ has a value between 0 and 1, the normalization constant 
$A$ in Eq.~(\ref{eq:lambda}) is written as $4/(T_{\rm max} - T_{\rm min})^2$ 
to make the maximum survival rate unity.

Once Eq.~(\ref{eq:rho_tr}) combined with Eqs.~(\ref{eq:lambda}) and
(\ref{eq:G}) yields the midge density at every site at time $t$, we need to
describe  how adult midges lay eggs in space. A plausible assumption is that adult
midges at ${\bf r}$ lay eggs on trees located not far from ${\bf r}$. 
Accordingly, the heterogeneity of the tree density must be taken into account
for the spread pattern of eggs. In our notation, adult midges at time $t$
are from eggs at time $t$ which were laid by 
adult midges at $t-1$. Accordingly, the egg density 
$\overline{\rho}_{t}$ at time $t$ must be related to 
midge density $\rho_{t-1}$ at time $t-1$, and we write the relation in the form:
\begin{equation}
\label{eq:rhobar}
\overline{\rho}_{t}({\bf r}) = \sum_{{\bf r}' \in {\cal N}({\bf r})} \rho_{t-1}
({\bf r}') \omega_{t-1}({\bf r}', {\bf r}),
\end{equation}
where ${\cal N}({\bf r})$ is the set of discrete lattice points within a
distance of 5km from ${\bf r}$ since the speed of midge spread has been 
reported to be about 5$\sim$6km/year~\cite{Park_thesis}. 
In Eq.~(\ref{eq:rhobar}), 
$\omega_{t-1}({\bf r}', {\bf r})$ controls how many eggs are laid at ${\bf r}$ 
by adult midges at ${\bf r}'$, and we assume the following form: 
\begin{equation}
\label{eq:omega}
\omega_t({\bf r}', {\bf r}) = g_m \frac{\psi_t ({\bf r}) }{ \sum_{{\bf r}'' \in {\cal N}({\bf r}')} \psi_t ({\bf r}'')},
\end{equation}
which implies that the adult midges at ${\bf r}'$ move into their local
neighbors and lay eggs in proportion to the local tree density.  
In this process, we assume that all midges have
the same reproductive capability; thus, the growth parameter of a midge, 
$g_m$ in Eq.~(\ref{eq:omega}), is suitably set to a uniform value of $5.0$ in our simulations. 

Various pest species including PNGMs need host vegetation for reproduction. 
Once invaded by parasites, hosts become weak due to the lack of 
water and nutrients and can face fatal situations in harsh circumstances.
The pine tree is the host vegetation for PNGMs, and once infected by PNGMs 
it loses leaves. Our equations, Eqs.~(\ref{eq:rho_tr})-(\ref{eq:omega}), so far deal
with how the midge density and egg density evolve in time. The last ingredient
in our model is to mimic the effect of midges on pine trees. 
In the absence of midges, the tree density increases at a constant rate every year until
the maximum possible density is approached. On the other hand, a large midge density
reduces the tree density.  The density $\psi_t({\bf r})$ of pine trees is thus
assumed to evolve in time as  
\begin{equation}
\label{eq:psi_tr}
\psi_t({\bf r}) = \min[(1+g_p) \psi_{t-1}({\bf r}) - \rho_t({\bf r}), \psi_{\rm max} ] , 
\end{equation}
where $g_p$ is the growth rate for pine trees per year and is set to 0.05 in the present work, 
and $\min(x,y) = x$ for $x < y$ and $\min(x,y) = y$ otherwise. 
The tree density in reality cannot grow indefinitely, which is reflected in the form
of the condition $\psi_t({\bf r}) \leq \psi_{\rm max}$. 
When the tree density at a location approaches the upper limit $\psi_{\rm max}$, which
is set to 1.0 in the present work, the tree density stops increasing. 

\subsection{Simulation Procedure}
\label{subsec:preset}

In summary of our model for the spatio-temporal evolution of PNGMs, we implement the
growth dynamics for the midge density in Eq.~(\ref{eq:rho_tr}) with the egg survival
rate in Eq.~(\ref{eq:lambda}) which depends on climate conditions like temperature
and precipitation. The previous generation of midges lay eggs, depending on local tree
density, as in Eq.~(\ref{eq:rhobar}), and the growth of trees is affected by 
the midge density as in Eq.~(\ref{eq:psi_tr}). Although our dynamics must be
a rough approximation of reality, we have tried to use the known results reported
in existing literature. One advantage of our model is that we can try various future
climate conditions through the use of the egg survival rate in Eq.~(\ref{eq:lambda}) 
which depends on temperature and precipitation.

The time evolution of the midge density, the egg density, and the tree density
are governed by the framework presented in Sec.~\ref{subsec:model}.
We use two future climate scenarios, RCP4.5 and RCP8.5, for this purpose 
and download the climate data from Climate Information Portal~\cite{CIP}.  
The original data have a temporal resolution of one month and a spatial resolution of 1km, 
	and we make average over time, as explained in Sec.~\ref{subsec:model}, to get 
the temperature $T_t({\bf r}$) and the precipitation $P_t({\bf r})$.  
In model simulations, we use a two-dimensional square lattice with a lattice
constant having a linear size 1km.

We use $\psi_{t=0}({\bf r}) = 1.0$ as an initial condition for the tree density.  
Of course, pine trees are not spread uniformly across South Korea, 
and we have more trees in mountainous areas. Only for simplicity, 
we use a uniform distribution of the tree density. 
We choose three harbor cities, Incheon, Mokpo, and Busan, and we assume that
the initial PNGM breakout spread starts from there. 
The reason is that the midge population 
enters Korea mostly through timber imported from abroad.
We assume that all midge eggs hatch to become midge worms, but only part of
the worm population becomes adult midges. Accordingly, we use the uniform initial 
condition $\rho_{t=0}({\bf r}) = 0$ everywhere, but the initial egg density
$\overline{\rho}_{t=0}({\bf r}_i) = 0.005$ is assigned at one single lattice point 
${\bf r}_i$ depending on which harbor city is the source position. 

Once initial conditions for $\rho$, $\overline{\rho}$ and $\psi$ are given with fixed
all parameter values, we first calculate the evolution of the midge density at
each site from Eq.~(\ref{eq:rho_tr}).  In this procedure, climate conditions
and the density of trees are used to calculate the midge density [see 
Eqs.~(\ref{eq:lambda}) and (\ref{eq:G})].
We then update the density of eggs and the density of trees by using
Eq.~(\ref{eq:rhobar}) and Eq.~(\ref{eq:psi_tr}), respectively. 
Adult midges are assumed to put their eggs following Eq.~(\ref{eq:omega}).
We assume that the initial breakout starts from one of the above mentioned
three harbor cities at year 2021 to investigate the spreading pattern in later years
till 2050.

\section{Results}
\label{sec:results}
\begin{figure*}
\includegraphics[width=0.95\textwidth]{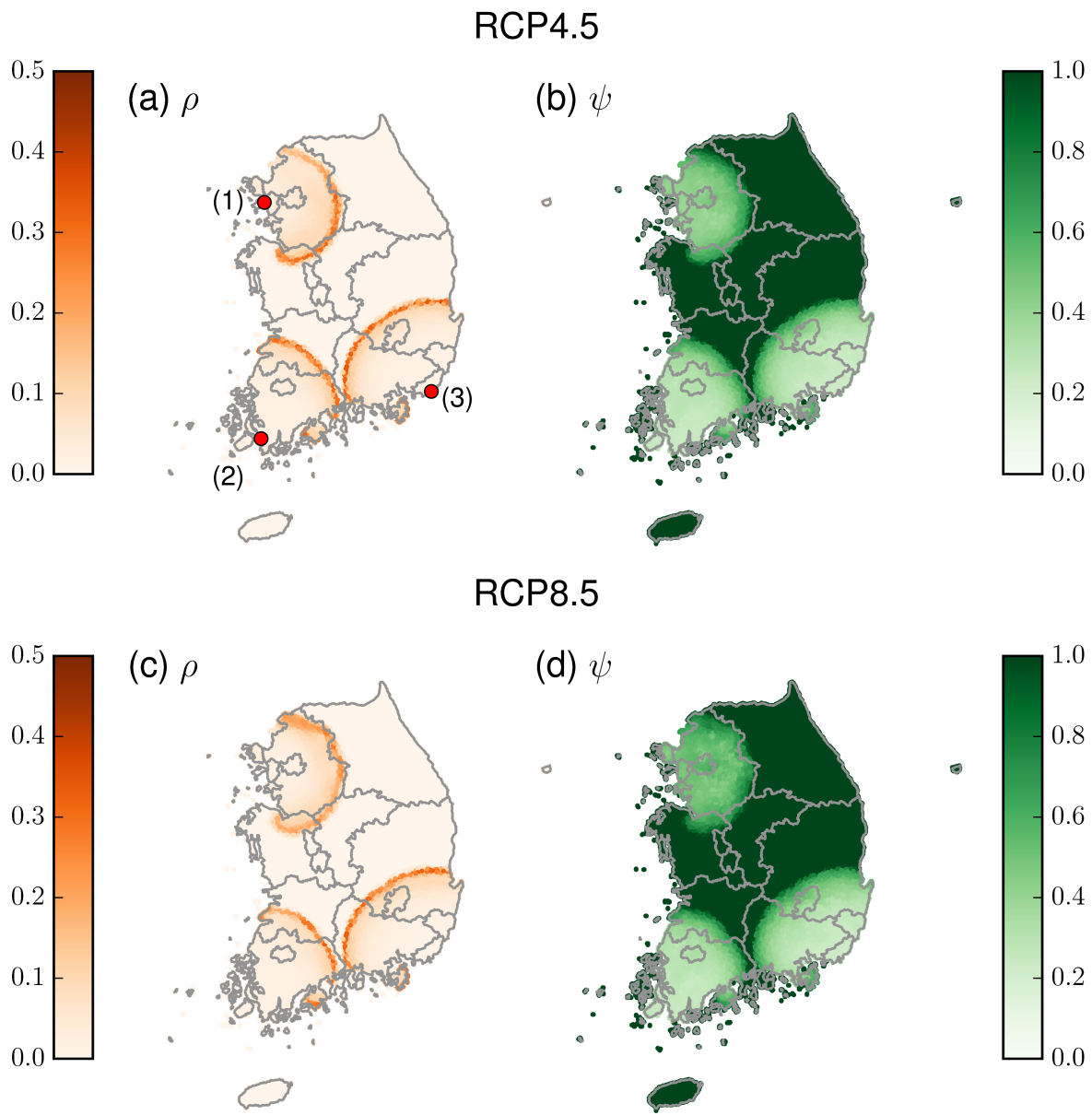}
\caption{The densities of PNGM ($\rho$) and trees ($\psi$) in $2050$ 
based on two climate conditions: RCP4.5 for (a) $\rho$ and (b) $\psi$, 
and RCP8.5 for (c) $\rho$ and (d) $\psi$, respectively. 
Initial breakout of PNGM is assumed to occur
in a harbor city [(1) Incheon, (2) Mokpo, or (3) Busan as shown in (a)] in 2021. 
Since spread of midges starting from one of three harbor cities is still
spatially limited in 2050, we plot three different results for three different
source locations in one Korean map for convenience, both for midge density and
for tree density.
The results from RCP4.5 and RCP8.5 are not much different from each other, but
the spread starting from Incheon is shown to be a little faster in RCP8.5 than in RCP4.5.
We note that the density of trees in the region where PNGM has already killed most trees 
remains small, but nonzero, in coming years.
}
\label{fig:PNGM2050}
\end{figure*}

In the 1920's, PNGMs began to spread in Korea. The source locations of spread
were harbor cities like Incheon, Mokpo, and Busan, where imported
pine timber were unloaded. If timber were infected by PNGM worms, 
the initial spread of PNGM could start from these harbor cities.
We first study how PNGMs would spread in the future if infected timber is imported
to one of the harbor cities in Korea. The key ingredient is the future climate
conditions, and we use the predictions based on the scenarios RCP4.5 and RCP8.5.
We emphasize that our main goal here is to study how future climate changes
can alter the pattern of parasite spread in general. Even though our model
parameters fit better for PNGM, the model can easily be generalized for
other parasite insects. 

In our simulations of spread, we assume that the midge spread starts in year
2021 and compute how the midge density evolves in time until the 2050.  For
simplicity, we pick $t=0$ for the year 2021; thus, 2050 corresponds to $t=29$.
For the initial condition of midge density, we use $\rho_0({\bf r}) = 0.0$ for
all locations; i.e., no adult midges exist at $t=0$.  For the initial values of
the egg density, we set $\overline{\rho}_0({\bf r}) = 0.005$ for one lattice point
in the harbor city area where the initial spread occurs. Of course, any location
,except for this source lattice point, is assigned $\overline{\rho}_0({\bf r}) = 0$.  
For the location of
the source of midge spread, we pick three cities, Incheon, Mokpo, and Busan
based on what happened in 1920's.  We then simulate our model for all six
($ = 3 \times 2$) different cases, i.e., three source locations and two climate conditions.

The spread of midges is shown to be not so fast and even after almost 30
years from the initial breakout of spread, midges are found not to have spread
across the country as shown in Fig.~\ref{fig:PNGM2050} where we display midge densities
in (a) and (c) together with tree densities in (b) and (d). 
For simplicity, we plot the results from three different
breakout locations (Incheon, Mokpo, and Busan) altogether in one Korean map
for a given future climate scenario RCP4.5 for (a) and (b), and RCP8.5 for (c) and (d). 
As is clearly seen,
the midge density propagates in space like a wavefront, and three density waves
starting from three locations do not meet yet. 
Our model dictates how midges tend to migrate to sites where
climate conditions are better and where more trees exist. 
The midge density propagates to locations far from the initial
breakout site as time goes on. 
Infected pine trees tend to die [see Eq.~(\ref{eq:psi_tr})] 
and it takes a long time for pine trees to recover their initial level of tree density.
Consequently, the locations where midge densities are high tend to form 
a circle-like structure, the radius of which expands in time. When pine trees
die of PNGM infection, PNGMs can hardly flourish later because of the lack of
the trees in which PNGM worms can survive. We thus expect PNGM density eventually to spread from
the initial source to the whole country after which they become extinct. Without
PNGMs, pine trees grow back and approach a suitable level of tree density. 
The circular shape of the wavefront of midge density originates from
the isotropy in the model. However, in reality, the pine trees do not
grow everywhere, and there are regions where pine trees grow better or worse exist.
We believe that our model has room for further improvement by making
the maximum tree density $\psi_{\rm max}$ depend on ${\bf r}$.

\begin{figure}
\includegraphics[width=0.45\textwidth]{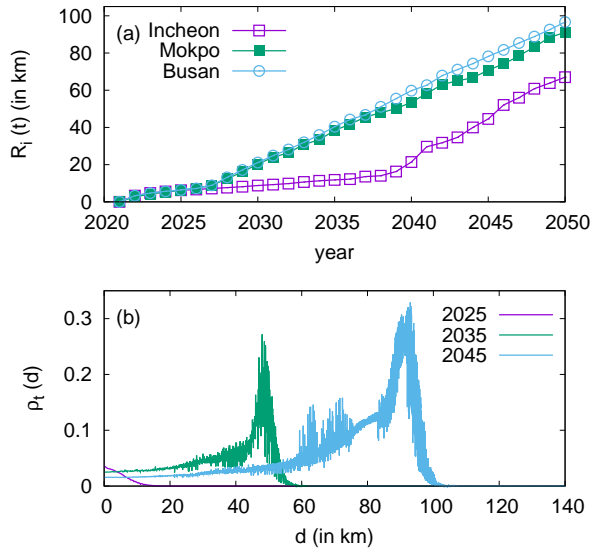}
\caption{(a) The radius of spread in Eq.~(\ref{eq:Rit}) increases as time
proceeds. Three harbor cities, Incheon, Mokpo, and Busan, are used as the
initial breakout sites of the PNGM spread, and the climate condition is based on
the RCP4.5 scenario.  After an early stage in which the radius increases very
slowly, the radius kicks off beyond the outbreak points which depend on source
cities. The radii of spread for Mokpo and Busan show sudden changes in the 
spreading speed at around 2026, but Incheon shows a much later outbreak point of
2038.  (b) The midge density profile at different times, 2025, 2035, and 2045,
for RCP4.5-based climate condition.  The distance $d$ from the source position with
Busan as the source city is used for the horizontal axis. 
}
\label{fig:radius_density}
\end{figure}

We define the spread radius $R_i(t)$ for the midge density when the spread has started
from the source city $i$ at position ${\bf r}_i$ as
\begin{equation}
\label{eq:Rit}
R_i(t) \equiv \sum_{\bf r} | {\bf r} - {\bf r}_i | \rho_t({\bf r}) . 
\end{equation}
For convenience, we choose ${\bf r}_i$ as one lattice point in
the harbor city $i \in \{${Incheon, Mokpo, Busan}$\}$, where the initial
condition $\overline{\rho}_0({\bf r}_i) = 0.005$ has been assigned.
At $t=0$, $\rho_0({\bf r})$ is
localized to this source location; thus, $R_i(t=0)  = 0$.  As time evolves,
$\rho_t({\bf r})$ extends to cover a broader region; thus, $R_i(t)$
increases.  

Figure~\ref{fig:radius_density}(a) displays the temporal change of $R_i(t)$ for
each source city when the climate scenario RCP4.5 is used.
In the initial stage of PNGM spread, the radius first increases very slowly, and 
then suddenly kicks off after some years.  
The radius increases at a rate of  1km/year for first five to six years
for Mokpo and Busan as a source locations. Similar stagnation is observed also for 
Incheon as source location, but it lasts for a much longer time until around the year 2038
with the slower increasing rate of 0.6km/year for the spread radius.
Interestingly a similar stagnation behavior has been
observed in reality for the past spread pattern of PNGMs~\cite{Park_thesis}.
In Ref.~\onlinecite{CA-pest-Chon}, the radius of spread was shown to
kick off after a stagnation period, and the outbreak point coincides well with the 
instant when the population approaches the carrying capacity. 
When the midge density is small,
the damage from parasites is soon repaired, which allows the midge to lay eggs
uniformly around its current position. In this case, the spatial midge density exhibits 
a unimodal shape with the center at the source location. As the midge population 
grows further, the ruined tree density cannot recover in a short time, but keeps decreasing 
gradually due to parasitizing midges. In such case of high midge density, the
diffusion of midges exhibits a bias toward an outgoing direction where more trees still exist.
Consequently, the location of the maximum midge density drifts away from the original
source location, and the radius of spread increases faster.  

When the midge spread starts from Incheon, the initial stage of stagnation 
is found to be longer than it is for the other source cites Mokpo and Busan. The difference
appears to originate from the different climate conditions, the precipitation
from March to May in particular. We observe that according to the RCP4.5-based
prediction, precipitation around Incheon gradually increases after year 2035
and thus $G(P)$ in Eq.~(\ref{eq:G}) takes the almost maximum value of unity at
years 2039 and 2040. This then leads to an increase in the midge density, which
soon reduces the tree density. If the tree density becomes smaller around the Incheon
area, the midges then migrate outward, and the radius in Fig.~\ref{fig:radius_density}(a)
kicks off at around year 2040.
After the early stage of migration, the locations where the midge densities are larger
begin to move out from the source position. When this happens,
the speed of diffusion is observed to become faster. The rate of increase of the radius
for Mokpo and Busan as source cites after the kickoff occurs are 
about 3.5km/year and 3.8km/year, respectively, while   
the corresponding value for Incheon is 4.7km/year.

In Fig.~\ref{fig:radius_density}(b), the density of midges with Busan as the
source city is shown as a function of the distance from Busan at three
different instants, 2025, 2035, and 2045. In the early stage of diffusion at
year 2025, the density shows a Gaussian-like shape with a maximum
at the position of the source city (at null distance) as expected.
After the early stage of diffusion, the Gaussian-like
shape with its maximum at the origin begins to change, and 
the maximum shifts away from the origin. This is due to the 
decrease in the tree density near the origin, which drives midges away in an outward direction, as explained above. 
Of course, if midges move away from the origin, the tree density
near the source location can recover. However, the midges do not come back to 
the origin 
because they have to overcome the harsh region in which the tree density is
lower. 
  

 \begin{figure}
\includegraphics[width=0.45\textwidth]{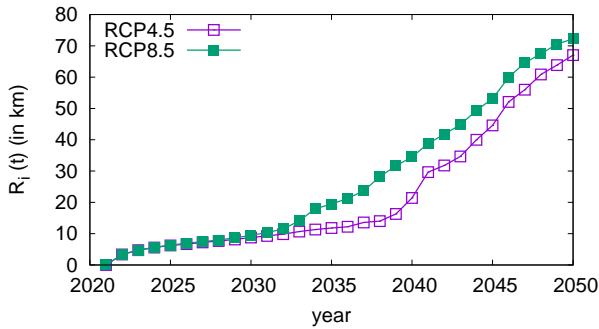}
\caption{The radius of spread $R_i(t)$ in Eq.~(\ref{eq:Rit}) for Incheon as the source
city with climate conditions based on RCP4.5 and RCP8.5. A sudden increase
in the radius of spread occurs earlier for RCP8.5 than for RCP4.5, which can
be explained by the sufficiently large value of precipitation in RCP8.5 (see
text for details).
}
\label{fig:RCP_45_85}
\end{figure}

We then investigate how different RCP scenarios affect the prediction of midge
density in the future. We use Incheon as initial source position and compare
the radius of spread obtained from RCP4.5 and RCP8.5 in
Fig.~\ref{fig:RCP_45_85}.  For RCP4.5, the radius kicks off at around $2039$	
while it kicks off at around $2031$ for RCP8.5. As explained above, the sudden
change in the slope in Fig.~\ref{fig:RCP_45_85} at these outbreak points originates
from the competition between two growth rates, that for trees and that for midges. The
difference in the outbreak points of RCP4.5 and RCP8.5 can  be explained as
follows: Precipitation significantly increases in the northern part of
South Korea as the greater greenhouse gas emission is assumed. 
Increased precipitation yields the increase of the midge population [see Eqs.~(\ref{eq:lambda}) and
(\ref{eq:G})], which shifts the outbreak point to an earlier time. We recognize
that the difference of such outbreak points between RCP4.5 and RCP8.5 is almost
indiscernible for other source cities, Mokpo and Busan.  It suggests that the
climate conditions in southern cities are already suitable for the fast growth
of the midge density.

\section{Conclusion}
We have proposed a spatio-temporal spread model of an insect species like the pine
needle gall midge (PNGM) that parasitizes trees. The model includes
the density of adult midges, the density of midge eggs, and the density
of pine trees as dynamic variables and describes how their dynamics
are coupled to each other. One of the main research goals of the present
paper has been to investigate how future climate conditions can alter the spread pattern
of insects. For this purpose, we have used climate predictions based on two standard
scenarios, RCP4.5 and RCP8.5, with different future estimates of
greenhouse gas emission. We have downloaded future climate data for each scenario
which contains grid data for temperature and precipitation with a temporal resolution of
one month. The density of PNGMs is calculated by the growth equation similar to 
the logistic equation, in which we use the climate-dependent survival rate.
The adult midges are modeled to lay eggs near their current positions in distance of
5km, and the density of eggs  depends on the local tree density.

From our extensive simulations, we observed that the radius of spread
as a function of time has different increase rates in the early and the late
stages of diffusion: In the early stage in which the midge density is still
small the diffusion of midges is slow whereas after the outbreak point of
spread the diffusion becomes much faster.  
We emphasize that a similar result of a
change in diffusion speed has been found in the field research~\cite{Park_thesis}. 
For a variety of different pests including PNGM investigated in
the present paper, the survival rate of parasites is greatly influenced by the climate
condition: More midges survive to become adults when both precipitation and
temperature are sufficiently high. 
We have  observed in our simulations that when Mokpo and Busan (located
along the southern cost of South Korea) are used as the source sites
of spread, the outbreak point occurs at a much earlier time than it does when Incheon
(located in the mid-western coast of Korean peninsula) is the source site.
We have investigated the reason for the difference between northern and
southern source sites and have found that it originates from the
different climate conditions, precipitation in particular. 
Southern regions of South Korea have sufficiently high precipitation
for midge population to grow fast, which then significantly reduces
the tree density near the source site. If this happens, midges tend
to migrate to a region with a high tree density and thus spread faster
in a radially outward direction from the source site. Consequently, the
outbreak point for the faster diffusion occurs in an early stage if the precipitation
near the source site is sufficiently high. 

As the greenhouse gas emission is increased, the RCP scenario
changes from RCP4.5 to RCP8.5, and climate variables such as temperature and
precipitation in the future are greatly influenced. 
Two different climate conditions, RCP4.5 and RCP8.5, 
have been found to result in almost the same spread pattern with a hardly recognizable
change of the outbreak points for Mokpo and Busan as source cities. 
In contrast, RCP8.5 yields a much earlier outbreak point than RCP4.5 for Incheon. 
We interpret that this difference in diffusion behavior between southern and northern source cites 
in South Korea should originate from the difference in the precipitation in the
two regions. In the southern part of South Korea, both RCP4.5 and RCP8.5 predict
sufficiently high precipitation and the spread kicks off early. However, for
the northern part of South Korea, the precipitation predicted by RCP8.5 is higher
than that predicted by RCP4.5, leading to a difference in the outbreak point for the 
case of Incheon as the source site. 
In more detail, an average precipitation higher than 
50mm has been observed to induce an increase in the midge density in successive $2 \sim 3$ years,
which reduces the host tree density; then, midges spread in an outward direction faster seeking 
regions with a high tree density. Such a condition of high precipitation near Incheon
has been found to be well satisfied at early times in RCP8.5, but not in RCP4.5, leading
to a shift in the outbreak point to an early time for RCP8.5.

We believe that our spatio-temporal growth model for PNGM spread
can easily be generalized for similar parasite insects. The present model
can also be generalized to mimic the spreading of parasites through road traffic.
For the case when infected timber is moved to other cities through ground
transportation, changing our model equation to cover such a
long-distance spread is straightforward. 

\section{Acknowledgment}
This study was carried out with the support of the Research Program of Rural Development Administration, Republic of Korea  (Project No. PJ01156304)

\end{document}